\documentclass{iaus}
\usepackage{graphicx}

\title[~~Panchromatic radiation from galaxies] 
{Panchromatic radiation from galaxies as a probe of galaxy formation and evolution}

\author[Michael Rowan-Robinson]   
{Michael Rowan-Robinson}

\affiliation{Astrophysics Group, Blackett Laboratory, Imperial College,
London SW7 2AZ}

\pubyear{2011} 
\volume{284}  
\pagerange{1--12}
\jname{The Spectral Energy Distribution of Galaxies}
\editors{R.J. Tuffs \&  C.C.Popescu, eds.}
\begin{document}

\maketitle

\begin{abstract}
I review work on modelling the infrared and submillimetre SEDs of galaxies.  The underlying physical assumptions are discussed and spherically symmetric, axisymmetric, and 3-dimensional radiative transfer codes are reviewed. Models for galaxies
with Spitzer IRS data and for galaxies in the Herschel-Hermes survey are discussed.  Searches for high redshift infrared and submillimetre
galaxies, the star formation history, the evolution of dust extinction, and constraints from source-counts, are briefly discussed.

\end{abstract}

\firstsection 
\section{Early work on radiative transfer models for infrared sources}

During the 1960s and early 1970s very simplistic models began to be developed for infrared
sources associated with circumstellar dust shells, HII regions and star-forming regions.  These 
models assumed that the dust grain density n(r) and temperature T(r) have a power-law dependence on 
radius r and that the dust is optically thin.  Scattering was neglected and
so the emission was simply calculated as a sum of  $n(r) Q_{\nu} B_{\nu} (T(r))$, integrated
over r.

The first complete solution of the spherically symmetric radiative transfer problem for an optically thick dust 
cloud was by  Leung (1975).
He used detailed grain properties and carried out a full solution of the moment equations, including 
anisotropic scattering.  The grain temperature was solved for by assuming radiative balance.
Yorke and Krugel (1977) modelled the dynamical evolution of an HII region, including the interaction between gas and dust.
Radiative transfer was approximated by diffusion.  These two papers marked a tremendous step forward in
the quality of models for infrared sources.  To take advantage of numerical techniques which had been used in
modelling stellar atmospheres, Leung essentially assumed that the radiation field is isotropic through most
of the cloud (the Eddington approximation).

Rowan-Robinson (1980) carried out a numerical solution of the full radiative transfer equation.  The intensity,
 $I_{\nu}(\theta,r)$, was calculated throughout the cloud using a grid of values in $\nu, r, \theta$.   This showed that
the Eddington approximation is a poor assumption throughout a dust cloud and revealed the phenomenon of sideways beaming of the radiation field at the inner edge of cloud.   Viewed from a distance a dust cloud surrounding a star
should show a ring of bright emission in the mid infrared, coming from the inner edge of the dust cloud.
A second interesting phenomenon revealed by a full solution of the radiative transfer equation is that
of back radiation onto the star.  The dust cloud radiates at the star and so the net spectrum of emission emerging
from the star is no longer a blackbody, but has absorption bands at the wavelengths of dust emission features
(Rowan-Robinson 1982).  Rowan-Robinson and Harris (1982, 1983ab) 
applied this code to all the bright circumstellar dust shells in the AFGL survey.



\section{Physical assumptions}

In modelling the infrared spectrum of star-forming regions and of galaxies, the key issues are the nature
of the dust grains and their spatial distribution.   A range of grain species and grain radii are needed
to account for the main observed features of the interstellar dust extinction profile with wavelength:
carbonaceous grains, silicates, and PAHs.  While the larger grains tend to be amorphous in most astrophysical 
environments, the smaller grains may be crystalline, and the very smallest grains, the PAHs, are essentially large 
molecules.  The possibility that the different grain species may be aggregated into larger fractal-like structures
has not been much explored.

The earliest radiative transfer models were targeted at either circumstellar dust shells or HII regions associated
with massive star formation.  For the latter the assumption of spherical symmetry is valid for the early,
compact phase of HII region evolution.  But many of the more evolved HII regions show a blister-like
geometry, as the HII region breaks out of the parent molecular cloud.  A starburst in a galaxy will involve
stars in both stages of development as well as in the later supernova phase.  Not many published
models attempt to capture this variety of geometries.  In ultraluminous infrared galaxies we are
almost certainly dealing with multiple star-forming clouds, so the heating of a cloud by its neighbours
can probably not be neglected and a full 3-D treatment is needed.

When we look at the wonderful images of the Milky Way produced by the Herschel mission, we can not
fail to be struck by the chiaroscuro distribution of the dust.  We see a complex alternation of optically
thick and optically thin regions, illuminated from many directions.   At the very least a clear distinction between
optically thick dust clouds and the optically thin intervening medium needs to be maintained in the
models.

\section{Spherically symmetric radiative transfer models for starbursts}

Despite their rather limited physical validity, it is not surprising that spherically symmetric radiative
transfer models have been widely deployed in the literature for massive star-forming regions (eg 
Krugel and Siebenmorgen 1994, Silva et al 1998).  
Efstathiou, Rowan-Robinson and, Siebenmorgen (2000) made such models slightly more realistic
by following the evolution of a spherically symmetric HII region within a dense molecular cloud.
They assume an embedded phase, lasting for $ 10^7$ years.  Following the star's explosion as a supernova,
the evolution continues as an expanding neutral shell, lasting from $10^7$ to $10^8$ years.
At $10^8$ years the emergent spectrum is virtually indistinguishable from that of the general
optically thin interstellar medium (the 'cirrus').  The time-sequence of SEDs calculated by
Efstathiou et al (2000) poses the interesting question: can we identify starbursts of different ages
observationally (see section 8 below) ?

 Subsequent work on radiative transfer models for starbursts includes the models of Takagi et al (2003), 
which includes distributed stars and dust satisfying a King law; the programme of
Dopita et al (2005, 2006ab), Groves et al (2008) which incorporate detailed HII region physics;
and the large library of parameterised models generated by Siebenmorgen and Krugel (2007).

\section{2-D, axially symmetric, radiative transfer models for infrared sources}

The problem of radiative transfer in an axially symmetric dust cloud was first solved by
Eftstathiou and Rowan-Robinson (1991) and applied by them to the problem of models
for protostars.  More recent axially symmetric models for protostars have been developed by
Whitney et al (2003, 2004).

Axially symmetric radiative transfer models for AGN dust tori were developed by
Pier and Krolik (1992), Granato and Danese (1994), and Efstathiou and Rowan-Robinson
(1995).  The main issue addressed by these papers was why we do not see the 10 $\mu$m
silicate feature strongly in emission from Type I QSOs.  Granato and Danese proposed to solve 
this by reducing the abundance of silicon, while Efstathiou and Rowan-Robinson suggest that
a tapered disk geometry shields the hot dust from our direct line of sight.

High quality mid infrared spectra from the Spitzer IRS for a large sample of starbursts and AGN
allow us to test these ideas.  Rowan-Robinson and Efstathiou (2009) have shown that the suite
of starburst models of Efstathiou et al (2000) and AGN dust tori models of Efstathiou and Rowan-Robinson
(1995) successfully explain the distribution of starbursts and AGN in the diagnostic diagram 
proposed by Spoon et al (2007),
9.7 $\mu$m silicate depth versus 6.2 $\mu$m PAH equivalent width (Fig 1).  While there is aliasing of the
models in this diagram, since different combinations of starburst and AGN dust torus models can
predict the same point in the diagram, the model sequences do populate the distribution of 
observed points well.   In particular the range of AGN dust tori silicate optical depths, from very deep
absorption to weak emission is well matched. It will be interesting to model the individual spectra to 
gain a more precise determination of the model parameters.

\begin{figure}[b]
\begin{center}
 \includegraphics[width=3.0in]{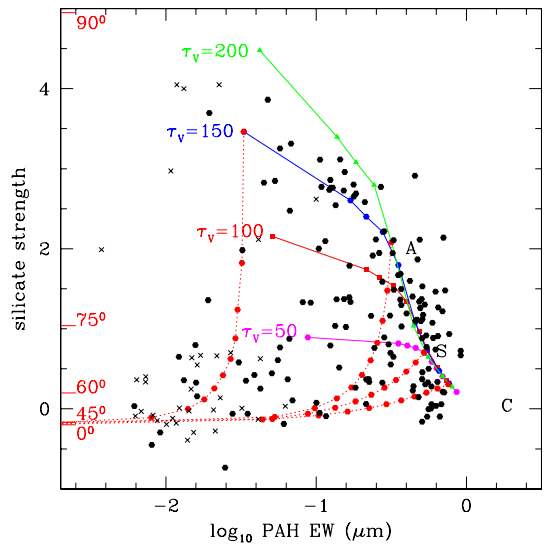} 
 \caption{9.7 $\mu$m silicate depth versus 6.2 $\mu$m PAH equivalent width for starbursts and AGN (Spoon et al 2007).
The observed points are well represented by the grid of models (Rowan-Robinson and Efstathiou 2009).}
   \label{fig1}
\end{center}
\end{figure}

\section{3-D radiative transfer models for clumpy AGN dust tori}
 
A first attempt to approximate a fully 3-dimensional radiative transfer was made by Rowan-Robinson (1995),
who argued that the dust torus around AGN might comprise a large number of small but optically
thick clouds.  The emission from the clouds was approximated by a slab model, and because these clouds
are found to be approximately isothermal, the absence of a 10 $\mu$m emission feature could be explained. 
Nenkova et al (2002, 2008), advocated a 5-10 clump model for NGC1068, also using a slab treatment of the 
radiative transfer.

Fully 3-dimensional radiative transfer codes based on Monte Carlo methods have been developed by
Dullemond and van Bemmel (2005), Shartmann et al (2005, 2008), and Hoenig et al (2006).   These codes have
the capacity to fully track the observed chiaroscuro of the dust distribution in galaxies and AGN.

\section{Radiative transfer models for quiescent disk galaxies ('cirrus')}

For optically thin dust in the interstellar medium of galaxies we only need to ensure thermal balance between
the grains and the interstellar radiation field.  The earliest radiative transfer model proposed for the 'cirrus' 
component in galaxies found by IRAS was given by Rowan-Robinson and Crawford (1989),
who modelled all IRAS galaxies detected in all four IRAS wavebands in terms of three components: an
M82-like starburst, an AGN dust torus and a cirrus component.  This cirrus model was extended by
Rowan-Robinson (1992), who found that the key parameter in characterising the cirrus spectrum was
the ratio of the intensity of the radiation field to that of the solar neighbourhood, $\psi$.  The shape of
the spectrum of the interstellar radiation field was of less importance.

Silva et al (1998) substantially improved this treatment by modelling the interstellar dust in galaxies
as a spheroidal distribution.  Regions of massive star formation were modelled with a spherically
symmetric code to generate a combined cirrus and starburst SED.  A chemical evolution code followed the
star formation rate, the gas fraction and metallicity, with the gas being divided between the star-forming
molecular clouds and the diffuse medium.  Further detailed models of spiral galaxies have been
developed by Popescu et al (2000, 2011), Misiratis et al (2001), Dale et al (2001), Tuffs et al (2004),
Piovan et al (2006).

Efstathiou and Rowan-Robinson (2003) developed a simple suite of cirrus models for local galaxies,
assuming an optically thin interstellar medium, characterised by the extinction $A_V$ (not $>>$ 1).
The radiation field was determined by Bruzual and Charlot (1993) starburst models,  parameterised by the age $t_*$, 
and an exponential decay time $\tau$ (for local galaxies, $t_*$ = 0.25 Gyr, $\tau$ = 5-11 Gyr).  Galaxies are
 then characterised by a single mean intensity, $\psi$  =  bolometric intensity$/$solar neighbourhood intensity.
They also found that cirrus models, with slightly higher $A_V$ and $\psi$ ( $\sim$ 2-3 times higher), 
can also fit high-z galaxies from the SCUBA blank-field surveys.  Restricting analysis to SCUBA sources: 
(a) which have been confirmed by submm interferometry, or (b) sources from 8 mJy survey which have radio 
associations, they found that 70$\%$ of submillimetre galaxies (16/23) were successfully modelled by cirrus model.  
More recently Efstathiou and Siebenmorgen (2009) have confirmed this result by modelling 12 submillimetre 
galaxies which have IRS spectroscopy.  They found that 3 were fitted with a pure cirrus model, the rest with 
a combined cirrus+starburst model.

\begin{figure}[b]
 \includegraphics[width=2.7in]{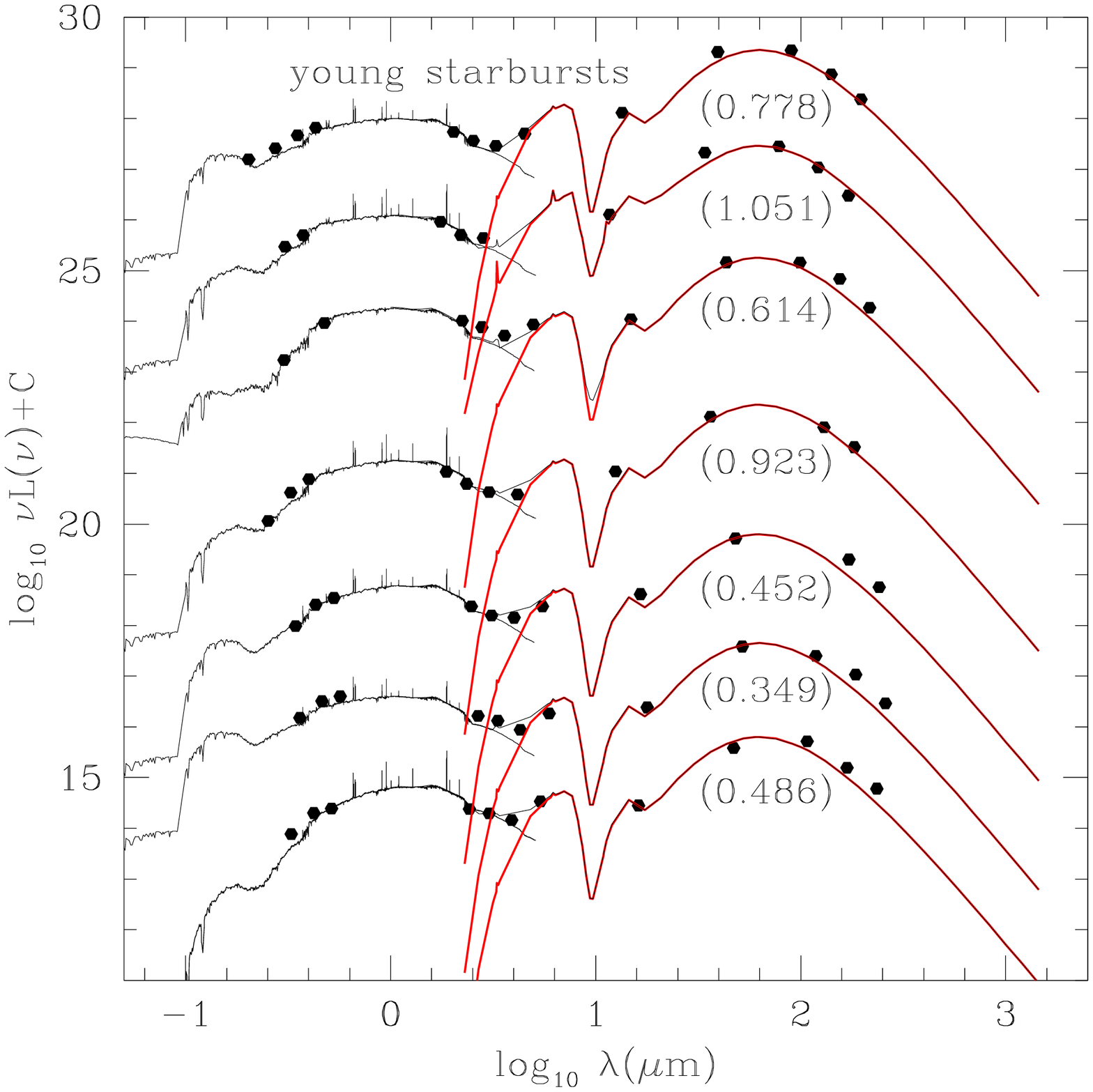} 
 \includegraphics[width=2.7in]{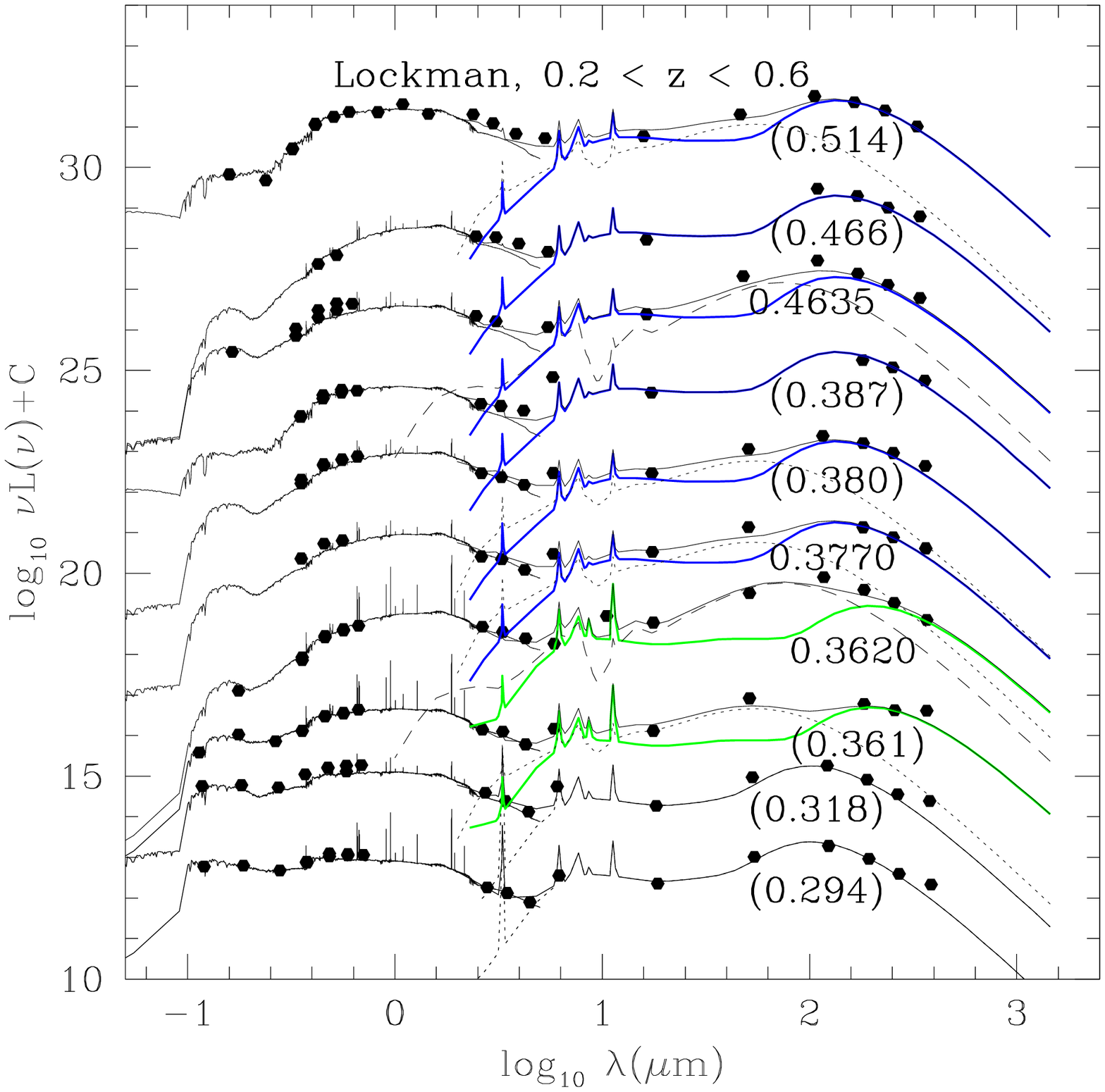}
 \caption{L: Young starbursts in the Hermes survey. R: Galaxies in the Hermes-Lockman area fitted with infrared
templates, including cirrus models with $\psi$ = 1(objects 5,6,8,10 from bottom) and 0.1 (objects 3, 4 from bottom).
Galaxies are labelled with their photometric redshift. From Rowan-Robinson et al (2010). }
   \label{fig2}
\end{figure}

\section{Infrared templates}

Several groups have generated suites of models for starbursts, AGN dust tori, and for
quiescent (cirrus) galaxies and these have been applied to modelling the infrared and submillimetre
SEDs of hundreds of thousands of galaxies.  Typically each component has two or three key parameters
(for starbursts, the initial optical depth and the age; for AGN dust tori, the inclination and torus
geometry; and for cirrus, $\psi$  and perhaps the SED of radiation field).  It is beyond the scope of 
this short review to try to review all these SED modelling efforts.   Instead I pick out a couple
of issues that have emerged from modelling galaxies detected by Herschel.

The first question is: are we beginning to be able to recognise young starbursts ?
Rowan-Robinson et al (2010) find a number of galaxies in the Herschel-Hermes survey
which seem to require a young starburst model (Fig 2L).   Secondly, how can we understand
the very cold dust that seems to be seen in some Herschel (and Planck) galaxies ?  Fig 2R
shows a number of relatively local galaxies which seem to require cirrus components with
$\psi = 1$.  This is not too startling since such galaxies were noted by Rowan-Robinson (1992)
in the IRAS data.  What is surprising is that there also seem to be some galaxies with $\psi = 0.1$.
If this really is optically thin dust, then it must come from extended dust distributions on scales
of tens of kiloparsecs.  Bendo (2012, this volume) argues that the cool dust seen in the outer regions
of nearby galaxies is heated by evolved stars not star formation. Such cool dust may be seen 
in dwarf galaxies (Madden 2012, this volume): an example is the dwarf galaxy NGC1705 studied by
O'Halloran et al (2010), for which they estimate a dust temperature of 5.8 K.

\section{ Evolution of galaxies - the infrared view: when was the first infrared light ?}

Turning to the evolution of galaxies over cosmological times, the search for high redshift (z$>$5) galaxies
is highly dependent on spectral synthesis models matched to near infrared photometry (Schaerer 2012,
this volume).  Leitherer (2012, this volume) has given a very nice review of issues in stellar spectral synthesis.

One further issue is that dust is a potential source of aliasing in Ôz$>$5Õ photometric redshifts, since 
extinction of radiation shortward of the Balmer break can mimic the Lyman break.   
Young stars forming today are emerging from $A_V \sim 100$ clouds.  What is the likely
dust optical depth in star-forming clouds at high redshift ?   We know that quasars at z$>$6 already have 
carbon, so these regions have already formed stars, and some have submillimetre emission so dust has formed.

When was the first (rest-frame) infrared light ?  Current ideas suggest that reionization took place between z = 11 and z = 6,  
the first stars formed at z $\sim$ 30, and the first  galaxies at z $\sim$ 10-20.  
How long did it take to get $A_V > 1$ in star-forming clouds and hence most starlight being reprocessed into the infrared ? 

If supernovae make enough dust to provide this opacity, then we need only $10^6$ years from the formation of the first
normal massive stars. However if we need AGB stars to make the dust, then we need at least  $10^9$ years. 
It is interesting to note that 1 Gyr from z=10 would be z=4, so redshift 4 is not an unreasonable estimate for the
moment when star formation starts to be universally a far infrared phenomenon.

In principle we can learn a lot about this era from dwarf galaxies, with metallicity extending down to 1$/$30th solar.  
The halo of our Galaxy also has stars with even lower metallicity, which could be from the z$>$10 era.


\begin{figure}[b]
\begin{center}
 \includegraphics[width=3.0in]{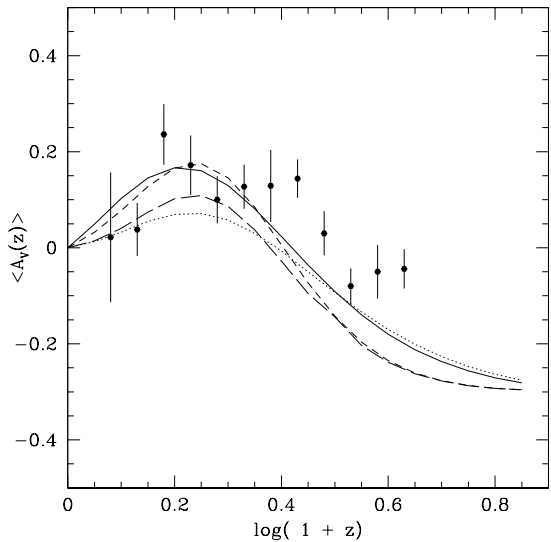} 
 \caption{Mean $A_V$ versus z for galaxies in HDFN (Rowan-Robinson 2003).
Curves are from closed box, instantaneous recycling, models for the evolution of gas
and dust in galaxies.}
   \label{fig3}
\end{center}
\end{figure}

\section{Submillimetre galaxies}

The first z $>$ 2 IRAS galaxy (IRAS F10214+4723) turned out to be bright at 850 $\mu$m (Rowan-Robinson et al 1991, 1993).
Franceschini et al (1994) pointed out that the negative K-correction at submillimetre wavelengths makes these a 
window to high redshift.  However, the observed redshift-distribution of submillimetre  galaxies peaks at 
z $\sim$ 2-3 (Chapman et  al 2005).  Source-count models seem to need  a redshift cutoff at $\sim$ 4 to explain 850 and 
1100 $\mu$m counts (see section 12).  The Herschel-Hermes survey detects plenty of z $>$ 3 galaxies 
(eg Rowan-Robinson et al 2010) and Herschel-ATLAS has detected a z = 4.2 starburst galaxy (Cox et al 2011).
But it still seems surprising that we have not detected more $z > 4$ submillimetre galaxies. 

Blain (1996) emphasised that gravitational lensing could be a big factor in explaining the high luminosities of 
submillimetre galaxies.  It is true that a significant fraction of IRAS hyperluminous infrared galaxies are lensed 
(Rowan-Robinson and Wang 2010 estimate the lensed percentage to be 10-30$\%$).
Negrello et al (2010) show 5 lensed galaxies from the Herschel-Atlas survey.
They estimate that a significant fraction (50$\%$?) of bright ($>$100mJy) 500 $\mu$m sources are lenses 
and predict that more than 100 lenses will be found.   The combination of the negative K-correction and gravitational
lensing should make it relatively easy to search for z = 4-6 infrared galaxies, if they are common.

\section{Star formation history}

Madau (1996) used the Lilley et al (1996) survey and data from the HDF to estimate star-formation rate as a function of redshift.
Rowan-Robinson et al (1997) used ISO data to derive much higher star-formation rates in the range z = 0-1
 ie star-formation rates derived from ultraviolet and optical data need a strong dust correction.
In using the far infrared luminosity to estimate star-formation rates it is important to correct for heating by 
evolved stars (Bell 2003, Rowan-Robinson 2003a).  With this correction, 
the far infrared luminosity is a good estimator for the obscured star formation rate.
A recent summary of estimates using different methods has been given by Hopkins (2007).  The mean
star-formation rate increases steeply from z = 0-1, peaks at z = 1 - 2.5 and then declines towards higher redshift.
This star-formation history can then be used to predict how the mean extinction in galaxies should 
vary with redshift (Rowan-Robinson 2003b, Rowan-Robinson et al 2008).  Fig 3 shows
observed points from a study of photometric redshifts in the Hubble Deep Field. 
The curves are closed box galaxy models with constant yield, instantaneous recycling, and a star-formation history 
consistent with the observed history, and in which all heavy elements are assumed to be in dust.

\begin{figure}[b]
 \includegraphics[width=2.7in]{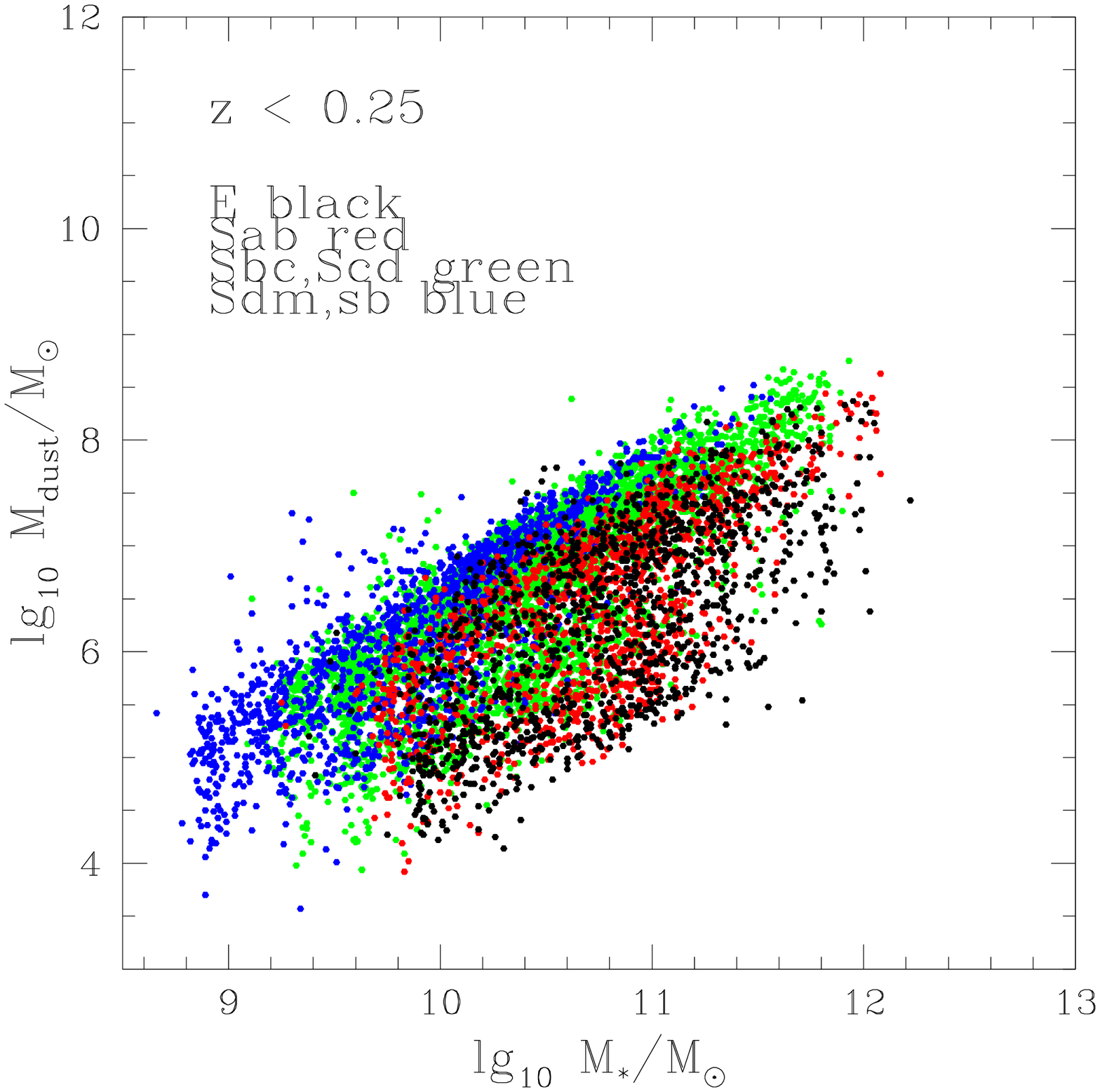} 
 \includegraphics[width=2.7in]{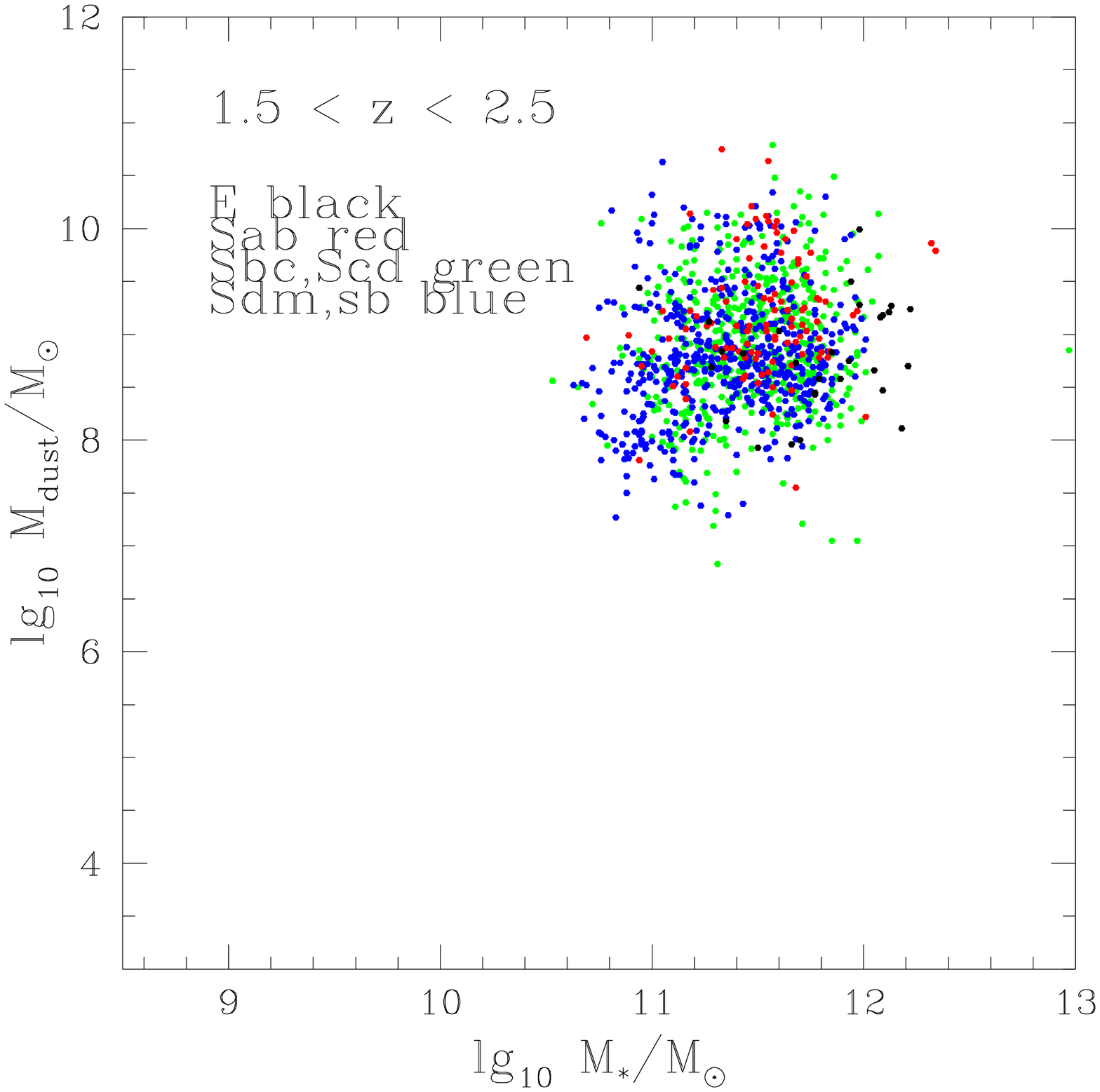}
 \caption{Dust mass versus stellar mass at z $<$ 0.25 and 1.5 $<$ 2.5 for galaxies in the
SWIRE survey (Rowan-Robinson et al 2008).  Galaxies detected at $z \sim 2$ in infrared surveys are
much dustier, and have much higher specific star formation rate, than their local counterparts.}
   \label{fig4}
\end{figure}

\section{Stellar mass, star formation rate, dust mass}

The importance of modelling infrared and submillimetre SEDs with radiative transfer models is
that key secondary parameters like star-formation rate and dust mass can be extracted.
From stellar spectral synthesis models for optical and near infrared SEDS, the stellar mass can be calculated.
For example the SWIRE Photometric Redshift Catalogue provides over 1 million redshifts (Rowan-Robinson et al 2008).
For most of these, stellar masses are estimated.  For those detected at 8 or 24 $\mu$m (25 $\%$ of the
catalogue), the star-formation rate and dust mass are estimated, as well as the luminosities in
each contributing component to the SED fit.  Recently a 
new version, based on improved and extended photometry, has been made available

(http://astro.ic.ac.uk/$\sim$mrr/swirephotzcat/zcatrev12ff2.dat.gz).

From stellar synthesis models, we can determine $M_* / L(3.6)$, and then use $L(3.6)$ to get $M_*$.
The evolution of $M_* / L(3.6)$ with time is approximated as  $\sim (a+b t^{-0.6})$, which is found to be accurate to 5$\%$.

For starbursts we can estimate the star-formation rate from $\phi_* (M_{\odot} / yr)$ = $2.2$ x $10^{-10} (L60 / L_{\odot})$  
(Rowan-Robinson 2001).  The specific star formation rate  $\phi_* /  M_*$   is then a measure of $t_{sf}^{-1}$, the inverse
of the time to generate the observed stellar mass, forming stars at the current observed rate.

From radiative transfer models for the infrared templates, we can estimate mass of dust, $M_{dust}$ in each component,
and sum this.   Figure 4 shows the  dust mass versus stellar mass, in two different redshift ranges.  Infrared galaxies
at $z \sim 2$ have both higher specific star-formation rates and dust masses compared to local galaxies.

\section{Far infrared and submillimetre source-counts}

An important way to constrain the star-formation history and AGN evolution history of galaxies is through
models for infrared and submillimetre source-counts and the integrated background radiation.  Recent 
source-count models which try to fit source-counts across the infrared and submillimetre band include those by
Rowan-Robinson (2009), Valiante et al (2009) and Bethermin et al (2011).  My models appear to require a
redshift cutoff at around $z \sim 4$ to be consistent with submillimetre counts and the background radiation.

Oliver et al (2010) have presented counts from the Herschel-Hermes survey at 250, 350, 500 $\mu$m 
and find that the counts, together with a P(D) analysis, account for a large fraction of infrared background.
Amblard et al (2011) have analysed clustering in the submillimetre background and conclude that submillimetre 
galaxies reside in dark matter halos with mass $ > 3$x$10^{11} M_{\odot}$.
This minimum mass scale for sub-mm galaxies corresponds to the most efficient dark matter mass scale for 
star-formation in the Universe.

\section{Conclusions}

(1) Modelling of ultraviolet to submillimetre SEDs has reached new levels of mathematical and physical sophistication.
The use of the simplistic models of the 1960s can not be justified.  At the very least the large published libraries 
of models calculated from radiative transfer codes should be used.

(2) Interesting cosmological insights can be expected from new missions (eg Herschel) and facilities (eg ALMA).

(3) It will be important to understand cold dust, and phenomena at low abundances.

(4) It may be hard work finding infrared galaxies at z $>$ 4 because only unusual environments have enough dust.

\end{document}